\documentstyle[12pt]{article}

 {\parindent 0.5cm \textwidth 15.0cm \textheight
23.5cm \hoffset=-0.8cm \voffset=-3cm
  \let\LARGE=\large
\let\Large=\normalsize \let\large=\normalsize
\newcommand{\be}{\begin{equation}}
\newcommand{\ee}{\end{equation}}
\newcommand{\ba}{\begin{array}{c}}
\newcommand{\ea}{\end{array}}
\newcommand{\dis}{\displaystyle}
\begin{document}
\begin{titlepage} \vspace{0.2in} \begin{flushright}
CPT-93/P.2920 \\ \end{flushright} \vspace*{1.5cm}
\begin{center} {\LARGE \bf  Chiral Defect Fermions and the Layered
Phase\\} \vspace*{0.8cm}
{\bf C.P. Korthals-Altes, S. Nicolis and J. Prades$^{a)}$}\\ \vspace*{1cm}
Centre de Physique Th\'eorique, C.N.R.S. - Luminy, Case 907 \\
F-13288 Marseille Cedex 9, France \\ \vspace*{1.8cm}
{\bf   Abstract  \\ } \end{center} \indent
 
Chiral defect fermions on the lattice in 4+1 dimensions are analyzed
using mean field theory. The fermion propagator has a localized chiral
mode in weak coupling but loses it when the coupling in the unphysical
fifth direction becomes too large. A layered phase {\it \`a la}
 Fu-Nielsen
appears where the theory is vector-like in every layer.
\vfill \begin{flushleft} June 1993 \\
CPT-93/P.2920 \vspace*{3cm} \\
\noindent{\rule[-.3cm]{5cm}{.02cm}} \\
\vspace*{0.2cm} \hspace*{0.5cm} ${}^{a)}$ 
Address after October 1st: NORDITA, Blegdamsvej 17\\
\hspace*{0.5cm} \hphantom{${}^{a)}$} DK-2100
Copenhagen \O, Denmark\end{flushleft} \end{titlepage}

\section{Introduction} \indent
 
A gauge invariant theory with chiral fermions is notoriously hard to
construct. There is a famous obstruction to having free chiral
fermions \cite{ks,nini} with a local, hermitian and
translation-invariant action on the lattice and many attempts to
circumvent it have been made (for a review cf. \cite{Rome}).
 
Recently appeared a proposal \cite{kaplan}, in which the theory is,
{\it a priori}, a vector theory in 4+1 dimensions. The mass of the
fermions depends on the  extra coordinate and and jumps from $-m$ to
$+m$. For sufficiently small gauge coupling and smooth gauge field
configurations, the presence of a chiral zero mode localized
at the jump was demonstrated \cite{CH}.  Its chiral
partner is pushed to the boundaries in the $x_5\equiv\xi$ direction.
Precisely how depends on the boundary conditions.
 
The idea now is to decouple all fermionic modes from the system except
for the chiral zero mode at the defect and be left there with a chiral
gauge theory. We will discuss this in the second Section.
 
In the next Section we will explore the phase diagram defined by the
physical gauge coupling $\beta$ and the ``unphysical'' gauge coupling
$\beta^{'}$. The latter couples the potential in the fifth
direction. Our technique will be mean field theory. For the pure
gauge case this was done by Fu and Nielsen~\cite{FN}, who discovered a
layered phase for $\beta^{'}$ sufficiently small and $\beta$
sufficiently large. In this phase subsequent four-dimensional
hyperplanes do not
interact and a mean field analysis shows that fermions are propagating
{\it inside } every hyperplane. The theory has become local in this
phase, hence the Nielsen-Ninomiya theorem strikes again: we are left
in each plane with a vector theory, thereby vetoing a proposal of
ref. \cite{kaplan} that a viable chiral theory results
at $\beta^{'}=0$.
 
In the last Section we discuss how to proceed further in the weak
coupling phase using mean field analysis.
 
\section{Weak coupling and chiral zero modes} \indent
 
In this Section we briefly review work done in
refs. \cite{kaplan,ucsd,ucsl}.
The system we study is a $U(1)$ theory coupled {\it vectorially} to
Wilson fermions
 
\be \renewcommand{\arraystretch}{1.2} \ba \label{sys}
\dis S(U,\Psi,\overline{\Psi})=\sum_{x}\left\{
\beta \sum_{\mu,\nu}^{4}\, \left(\makebox{\bf 1}-{\rm
Re}\, U_{\mu\nu}(x)\right)\,+\,\beta^{'}\, \sum_{\mu=1}^{4}\,
\left(\makebox{\bf 1}-{\rm Re}\, U_{\mu 5}(x)\right) \right.\\
\dis \hphantom{S(U,\Psi,\overline{\Psi})=\sum_{x}\left\{\right\}} \left.
+\,\sum_{\mu=1}^{5}\, \overline{\Psi}(x)\, \left[
\frac{\gamma_{\mu}}{2}\, \left( U_{\mu}(x)\, \Psi(x+\hat{\mu})
\, -\, U_{\mu}^{\dag}(x-\hat{\mu})\Psi(x-\hat{\mu})
                      \right)\right. \right. \\
\dis \hphantom{S(U,\Psi,\overline{\Psi})=\sum_{x}\left\{\right\}
+\,\sum_{\mu=1}^{5}\, }  \left.
\left.+\frac{r}{2}\left(U_{\mu}(x)\Psi(x+\hat{\mu})\,-\,
2\, \Psi(x)
U_{\mu}^{\dag}(x-\hat{\mu})\Psi(x-\hat{\mu})\right)
                                     \right] \right.\\
\dis \hphantom{S(U,\Psi,\overline{\Psi})=\sum_{x}\left\{\right\}
+\,\sum_{\mu=1}^{5}\, \overline{\Psi}(x)\, \left[\right]} \left.
+\, M(\xi)\, \overline{\Psi}(x)\, \Psi(x) 
\vphantom{\beta \sum_{\mu,\nu}^{4}\,} \right\} \, .
\ea \renewcommand{\arraystretch}{1} \ee
 
\noindent and we have that
 
\be \dis 
M(\xi)\equiv m\,{\rm sign}\left(\xi+\frac{1}{2}\right) \, .
\ee
 
\noindent
Our notation is standard. Note that the dimensionless Dirac mass is in
lattice units. Let us look at the Dirac equation following
from (\ref{sys}), setting for the moment $U_{\mu}(x)=\makebox{\bf 1}$.
It reads
 
\be \renewcommand{\arraystretch}{1.2} \ba \label{Dirac} \dis
\sum_{\mu=1}^{5}\left\{ \frac{\gamma_{\mu}}{2}\,
\left(\Psi(x+\hat{\mu})\, -\,
\Psi(x-\hat{\mu})\right)\, +\, \frac{r}{2}\left(\Psi(x+\hat{\mu})
\, -\, 2\,\Psi(x)\, +\, \Psi(x-\hat{\mu})\right)\right\}\\ \dis
\hphantom{\sum_{\mu=1}^{5}\left\{ \frac{\gamma_{\mu}}{2}\,
\left(\Psi(x+\hat{\mu})\, -\,
\Psi(x-\hat{\mu})\right)\, +\right\}}
+\, M(\xi)\, \Psi(x)\,=\,\lambda\,\Psi(x)\, .
\ea \renewcommand{\arraystretch}{1} \ee
 
\noindent
We search for a normalizable zero mode $(\lambda=0)$
\cite{rebbi}.
Because of translation invariance in the first four directions, we can
analyze eq.~(\ref{Dirac}) in terms of dimensionless lattice momenta
$-\pi\leq p_{\mu}<\pi,\,\,\mu=1,2,3,4$. The result, after some trivial
reshuffling,  is
 
\be \ba \renewcommand{\arraystretch}{1.2} \label{Dirac1} \dis 
\sum_{\mu=1}^{4}\, {\rm i}\, \gamma_{\mu}\sin p_{\mu}
\, \Psi(p,\xi) \,+\, \frac{\gamma_{5}+r}{2} \,
                    \Psi(p,\xi+1)\,+\,     \frac{-\gamma_{5}+r}{2}
                                        \,   \Psi(p,\xi-1) \,=\\
\dis 
\hphantom{\sum_{\mu=1}^{4}{\rm i}\, \gamma_{\mu}\sin p_{\mu}
\, \Psi(p,\xi) \,+\, \frac{\gamma_{5}+r}{2} }
\left(r+r\sum_{\mu=1}^{4}\left(1-\cos
p_{\mu}\right)-M(\xi)\right)\Psi(p,\xi) \, .
\ea \renewcommand{\arraystretch}{1} \ee
 
\noindent
For $r=1$, eq. (\ref{Dirac1}) separates into positive and negative
helicities and, for any doubler momentum $p^{D}$ (with $p_\mu^{D}=0\,
\,{\rm or} \,\pi$), we get
 
\be \ba \label{double} \dis
\Psi_{R,L}(p,\xi\pm 1)\,=\,\left(1+\sum_{\mu=1}^{4}\left(1-\cos
p_{\mu}^{D}\right)-M(\xi)\right)\Psi_{R,L}(p,\xi) \, 
\ea \ee
 
\noindent
where $\displaystyle\Psi_{R,L}\equiv\frac{1\pm\gamma_{5}}{2}\,
\Psi$. Let us analyze the right-handed case.
For $\xi>0$ the factor multiplying $\Psi_{R}(x)$ on the right-hand
side of eq. (\ref{double}) should be smaller than 1 in absolute value
 
\be \ba \label{norm} \dis
\sum_{\mu=1}^{4}\left(1-\cos
p_{\mu}^{D}\right)<m<2+\sum_{\mu=1}^{4}\left(1-\cos p_{\mu}^{D}
\right) \, .
\ea \ee
 
\noindent
Equation (\ref{norm}) tells us that the ``doubler''
$p^{D}=(0,0,0,0)$ has a
normalizable right-handed zero mode only for $0<m<2$.
For $2<m<4$ there are four normalizable {\it left-handed} zero modes,
corresponding to the four doublers $p^{D}=(\pi,0,0,0)$, etc.
 
In conclusion, eq. (\ref{norm}) says that for any fixed values of $m$
there is a definite number of normalizable zero modes with
well-defined chirality.
 
The zero mode becomes visible as a pole in the propagator of
the fermion, as shown by authors of ref. \cite{na-neu}.
 Taking as example $p^{D}=(0,0,0,0)$ and a physical
momentum $\bar{p}$ so that $p_{\mu}\equiv
a\bar{p}_{\mu}\to 0$, one finds a
right-handed pole localized around $\xi=0$ in the fermion propagator
$G(p)_{\xi,\xi^{'}}$. The above
analysis is valid for $U_{\mu}=\makebox{\bf 1}$, i.e.
$\beta=\beta^{'}=\infty$. In
the next Section we will see that the propagator behaves in a
radically different fashion when $\beta^{'}\to 0$.
 
\section{Mean field approximation} \indent
 
In this Section we analyze the physics of eq. (\ref{sys}) by mean
field techniques \cite{drouffe}. To this end we first fix the gauge
$U_{4}(x)\equiv\makebox{\bf 1}$ everywhere.
Then we write the free energy as a path integral
 
\be \label{free} \dis
e^{-F} = \int {\cal D} \overline{\Psi} {\cal D} \Psi {\cal D}
U \exp-S(U,\Psi,\overline{\Psi}) \ee
 
\noindent
and gauge fixing is tacitly understood.
 
The next step is to introduce complex variables $v_{\mu}(x)$ by
 
\be \label{vee} \dis
\int {\cal D} v \prod_{x,\, \mu\neq 4}\delta(v_{\mu}(x)-U_{\mu}(x)) = 1
\ee
 
\noindent
and the corresponding ``driving field'' $\alpha_{\mu}(x)$ through the
integral representation of the $\delta$-function
 
\be \renewcommand{\arraystretch}{1.2}
\ba \label{integral} \dis
1 = \int {\cal D} v\, \int_{-{\rm i}\infty}^{{\rm i}\infty}
\, {\cal D} \alpha \,  \exp \, \left(-\,\sum_{x,\, \mu\neq
4}\, \alpha_{\mu}(x)\, \left(v_{\mu}(x)\,-\, U_{\mu}(x)
\right)\right) \, .
\ea \renewcommand{\arraystretch}{1} \ee
 
\noindent
Inserting eq. (\ref{integral}) into eq.(\ref{free}) decouples the
integrals over the $U_{\mu}$. Carrying out that integration one finds
 
\be \ba \renewcommand{\arraystretch}{1.2} \label{effective} \dis
e^{-F} = \int {\cal D} \overline{\Psi} {\cal D} \Psi {\cal D}
v {\cal D} \alpha \,
\exp \, -\, S_{eff}(v,\alpha,\Psi,\overline{\Psi})
\ea \renewcommand{\arraystretch}{1} \ee
 
\noindent with
 
\be \renewcommand{\arraystretch}{1.2} \ba \label{effact} \dis
S_{eff}(v,\alpha,\Psi,\overline{\Psi}) \, = \, \sum_{x}\left[
\beta \, \sum_{\mu,\nu}^{4}
                              \left(\makebox{\bf 1}\, -\,
{\rm Re}\,v_{\mu\nu}(x) \right)\,+\,
\beta^{'}\, \sum_{\mu=1}^{4}\, \left(\makebox{\bf 1}\,-\,
{\rm Re}\,v_{\mu 5}(x)\right)\right.\\ \dis
\hphantom{S_{eff}(v,\alpha,\Psi,\overline{\Psi}) } \left.
+\, \sum_{\mu=1}^{5}\, \overline{\Psi}(x)\, \left\{
\frac{\gamma_{\mu}+r}{2}\, v_{\mu}(x)\Psi(x+\hat{\mu})
\right. \right.\\ \dis \hphantom{S_{eff}(v,\alpha,\Psi,
\overline{\Psi}) \,
= } \left. \left. +\, \frac{-\gamma_{\mu}+r}{2}\,
v_{\mu}^{\dag}(x-\hat{\mu})\, \Psi(x-\hat{\mu})
\, +\, (-5r+M(\xi))\Psi(x)\right\} \vphantom{\beta \, 
\sum_{\mu,\nu}^{4}} \right] \, .
\ea \renewcommand{\arraystretch}{1} \ee
 
\noindent
We will be interested in varying $\beta$ and $\beta^{'}$ independently
and search for a saddle point in order to evaluate
eq.~(\ref{effective}). The mean field approximation consists in
setting
 
\be \renewcommand{\arraystretch}{1.2} \ba \label{Ansatz} \dis
\langle U_{\mu}(x)\rangle = v(\xi) \hspace*{1cm} 
{\rm for }
\hspace*{1cm} \mu=1,2,3 \\ {\rm and} \\ \dis 
\langle U_{5}(x)\rangle = v^{'}(\xi);
\ea \dis \ee

\vspace*{0.5cm} 

\noindent
due to four-dimensional
 hypercubic invariance. Moreover, we will assume the saddle
point is real. Under these conditions the saddle point equations read
 
\be \renewcommand{\arraystretch}{1.2} \ba \label{saddle1} \dis
4\, \beta \, v^{3}(\xi)\,+\,2\, \beta v(\xi)\, +\,
\beta'\, \left(v(\xi+1)\, {v^{'}}^{2}(\xi)\, +\,
v(\xi-1)\, {v^{'}}^{2}(\xi-1)\right)\, +\, j_{\mu}(\xi) \,
= \, \alpha(\xi) \\ {\rm and} \\ \dis
6\, \beta^{'}\, v^{'}(\xi)\, v(\xi)\,
v(\xi+1)+2\beta^{'}v^{'}(\xi)\, +\, j_{5}(\xi) \, = \,
\alpha^{'}(\xi)
\, . \ea \renewcommand{\arraystretch}{1} \ee
 
\noindent
The relation between the mean field and the ``driving field''
is:
 
\be \renewcommand{\arraystretch}{1.4} \ba \label{saddle2}
\dis 
\frac{I_{1}\left(\alpha (\xi)\right)}{I_{0}\left
(\alpha(\xi)\right)} = v(\xi) \\ {\rm and} \\ \dis
\frac{I_{1}(\alpha^{'}(\xi))}{I_{0}\left(\alpha^{'}(\xi)
\right)} = v^{'}(\xi)
\ea \renewcommand{\arraystretch}{1} \ee
 
\noindent with
 
\be \label{saddle3} \dis
I_{k}(\alpha)\, =\, \int_{-\pi}^{\pi}\frac{d\phi}{2\pi}(\cos
\phi)^{k}e^{\alpha\cos\phi}\,.
\ee
 
\vspace*{0.5cm}

\noindent
The fermionic current terms $j_{\mu}(\xi)\,\,\,\mu=1,\ldots,5$ are
short-hand for
 
\be \renewcommand{\arraystretch}{1.2} \ba \label{fercur} \dis
j_{\mu}(x)\equiv \left\langle\overline{\Psi}(x)\frac{\gamma_{\mu}+r}{2}
\Psi(x+\hat{\mu})\right\rangle \,+\,
\left\langle\overline{\Psi}(x+\hat{\mu})
\frac{-\gamma_{\mu}+r}{2}\Psi(x)\right\rangle
\ea \renewcommand{\arraystretch}{1} \ee
 
\vspace*{0.5cm}

\noindent
and they are evaluated using the self-consistent fermion propagator
 
\be
G(x,x^{'})\equiv\langle\overline{\Psi}(x)\Psi(x^{'})\rangle
\ee
 
\noindent with
 
\be \renewcommand{\arraystretch}{1.3} \ba \label{prop} \dis
G^{-1}(x,x^{'})\,=\,\sum_{\mu=1}^{4}\,
\left\{v(\xi)\frac{\gamma_{\mu}+r}{2}
\delta_{\xi,\xi^{'}}\, \delta_{\bar{x}+\hat{\mu},{\bar{x}}^{'}}^{(4)}\,+\,
v(\xi)\frac{-\gamma_{\mu}+r}{2}\,
\delta_{\xi,\xi^{'}}\,\delta_{\bar{x}-\hat{\mu},{\bar{x}}^{'}}^{(4)}
\right\} \\ \dis 
\hphantom{G^{-1}(x,x^{'})\,=\,\sum_{\mu=1}^{4}\,}\,+\,
v^{'}(\xi)\,\frac{\gamma_{5}+r}{2}\,\delta_{\xi+1,\xi^{'}}\,
\delta_{\bar{x},{\bar{x}}^{'}}^{(4)}\,+\,v^{'}(\xi-1)\,
\frac{-\gamma_{5}+r}{2}\delta_{\xi-1,\xi}
\delta_{\bar{x},{\bar{x}}^{'}}^{(4)}\\ \dis
\hphantom{G^{-1}(x,x^{'})\,=\,\sum_{\mu=1}^{4}\,}
+\,(-5r+M(\xi))\delta_{x,x^{'}}^{(5)}
\ea \renewcommand{\arraystretch}{1} \ee
 
\noindent
where we have used the notation $x\equiv(\bar{x},\xi)$ and
eq. (\ref{prop}) incorporates the {\sl Ansatz} in
eq. (\ref{Ansatz}).
 
We see that, for large $\beta$ and $\beta^{'}$, the saddle point
equations (\ref{saddle1}) and (\ref{saddle2}) are insensitive to the
presence of fermions, leading to the solution $v=v^{'}=1$. There is
also a strong coupling phase, with $v=v^{'}=0$.
Most
interestingly, one also finds a solution with $v\neq 0$, $v^{'}=0$;
this is a
layered phase, as in the pure gauge case \cite{FN}, since $v^{'}=0$
renders the fermion propagator $G$ local in $\xi$. Hence the matrix
element $j_{5}(\xi)$ vanishes, consistent with $v^{'}=0$ from
eqs.~(\ref{saddle1}) and (\ref{saddle2})--there is a layered phase with a
fermion propagator that is strictly confined to four-dimensional layers.
This propagator is {\it vectorial}; and in the large $\beta$ limit (so
$v=1$) we just have the four-dimensional Wilson propagator
 
\be \renewcommand{\arraystretch}{1.2} \ba \label{wilson} \dis
G^{-1}(p;\xi,\xi^{'})\,=\,\left(
{\rm i}\sum_{\mu=1}^{4}\gamma_{\mu}\sin p_{\mu}\,+\,
r\,\sum_{\mu=1}^{4}\, 
\left(\cos p_{\mu}\,-\,1\right)\,-\,r\,+\,M(\xi) \right)
\,\delta_{\xi,\xi^{'}} \, .
\ea \renewcommand{\arraystretch}{1} \ee
 
\section{Summary and prospects} \indent
 
Clearly the fermion propagator changes dramatically from weak
coupling, i.e. $\beta,\beta^{'}\to\infty$, where $v=v^{'}=1$, to
$\beta^{'}$ small enough (but $\beta$ large enough), where $v\neq 0$
and
$v^{'}=0$. The localized chiral pole at weak coupling disappears
entirely when $\beta^{'}$ is small.
 
Intuitively this is perfectly understandable on two counts. First: the
coupling in the fifth direction becomes so strong ($\beta^{'}$ small)
that the chiral mode at $\xi=0$ starts to bind with its chiral partner
at the boundary of  the system. Second: the chiral mode at $\xi=0$ has
its chirality related to the gradient of the mass profile. In the
layered phase it is unable to detect the gradient and the bias for a
given chirality disappears.
 
What is small enough for $\beta^{'}$? In the absence of fermions,
a simple argument \cite{CH} shows that $\beta_{c}^{'}=1/4$. Including
fermions raises this value. For $\beta$ one finds, in the absence of
fermions, $\beta_{c}\approx 1$; this holds in the presence of fermions
as well. In the mass range that admits chiral fermions in the weak
coupling phase, we find shifts of the order of $10\%$ (cf.
Figure 1).  Note that the dotted line separating the phases $B$ and $C$
when fermions are included
is only indicative of what happens close to the triple point; while it
can be shown that its slope close to this point is negative, we
haven't mapped it out completely.
 
What one must do is to
unravel the surface effect, proportional
to $L_{\perp}^{4}$ (where $L_{\perp}$ is the size of the system in the
4 physical directions) as proposed in ref. \cite{na-neu}. 
To this end, take the free energy of the
system  with a constant positive (negative) mass, i.e.
$F_{+}\,(F_{-})$.  On the other hand,
the free energy, $F_{+-}$, of the system with the mass profile is
 
\be \label{fpm} \dis
F_{+-}\,=\,L\,L_{\perp}^{4}f_{+-}+L_{\perp}^{4}\sigma_{+-}+\cdots
\ee
 
\noindent
Since the fermion is heavy
$f_{+-}$ is additive so that
 
\be \label{fpmadd} \dis
f_{+-}\,=\,\frac{f_{+}+f_{-}}{2} \, .
\ee
 
\noindent Therefore the quantity
 
\be \label{fadv} \dis
F_{+-}\,-\, \frac{F_{+}\,+\,F_{-}}{2}
\ee
 
\vspace*{0.5cm}

\noindent
 gives the excess in free energy of
the defect. This free energy excess should be due to a chiral theory;
as such it will be the gauge field average of the effective action
where the gauge invariant part is due to the $\eta$ invariant
\cite{luis}.
This is under investigation.
 
\section*{Acknowledgements} \indent
 
C.P.K.-A. thanks the theory division of KEK (Tsukuba, Japan)
for its warm
hospitality and S. Aoki, G. 't Hooft, T. Kashiwa, Y. Kikugawa, H.
B. Nielsen, N. Ninomiya, M. Okawa and J. Smit for useful discussions.
The work of J.P. has been supported in part by CICYT, Spain, under Grant
No AEN90-0040. He is also indebted to the Spanish Ministerio
de Educaci\'on y Ciencia for  a fellowship.

\newpage  \pagestyle{empty} 
\begin{center} \section*{Figure Caption} \end{center}

\vspace*{2cm}

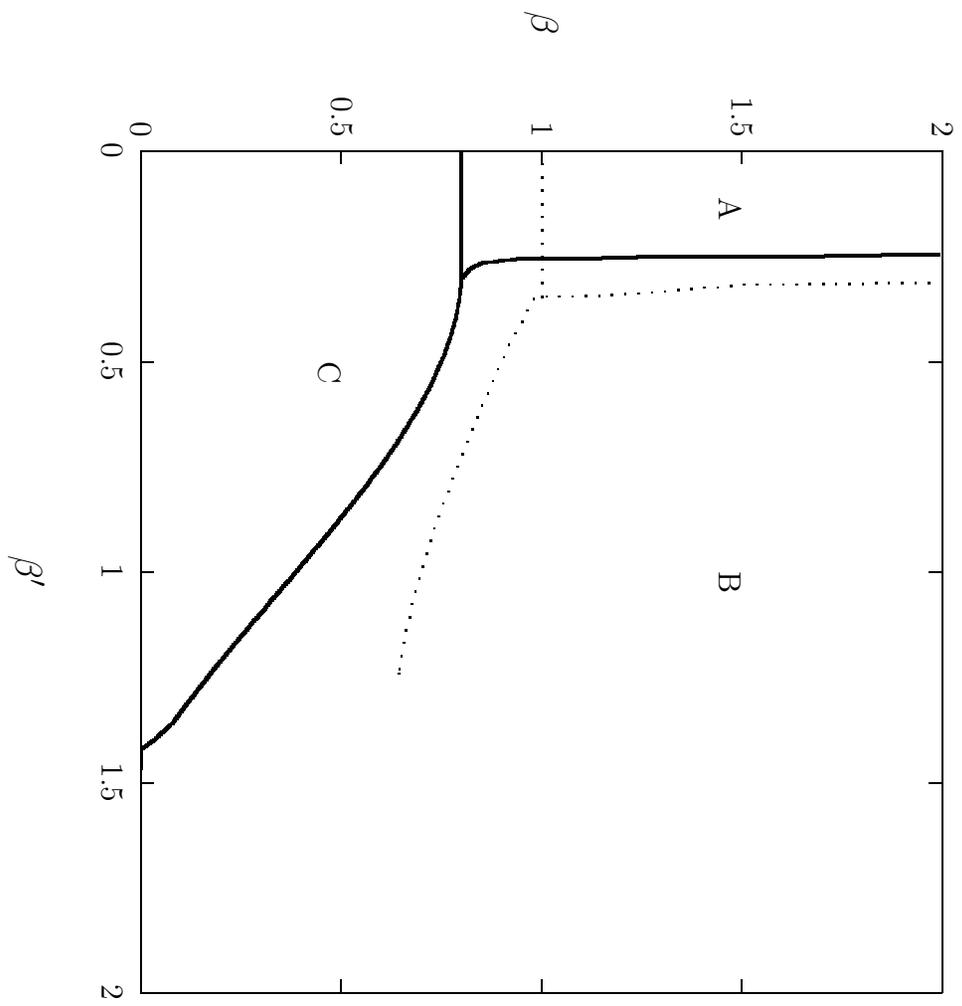
\noindent {\bf Figure 1}: \hspace*{1cm} 
 Phase diagram of the model. Solid (dotted) 
lines are for the absence (presence) of fermions. The three phases present
are: a layered phase ($A$), a weak coupling phase ($B$) and a 
strong coupling phase ($C$). 
}
\end{document}